\documentclass[prb,english,twocolumn,amsmath,amssymb,superscriptaddress]{revtex4-2}
\usepackage{amsmath}
\usepackage{graphicx}
\usepackage{amssymb}
\usepackage{dsfont}
\usepackage{color}
\usepackage{lipsum,kantlipsum}
\usepackage[bookmarks=true,colorlinks,linkcolor=blue,urlcolor=blue,citecolor=blue]{hyperref}

\usepackage{hyperref}
\usepackage{tikz}
\usetikzlibrary{calc}

\newcommand{\be}{\begin{equation}}
\newcommand{\ee}{\end{equation}}
\newcommand{\bea}{\begin{eqnarray}}
\newcommand{\eea}{\end{eqnarray}}
\newcommand{\ket}{\rangle}
\newcommand{\bra}{\langle}

\definecolor{darkorange}{rgb}{1, 0.55, 0.0}

\begin{document}
	
\title{Power-law decay of correlations after a global quench in the massive XXZ chain}

	\author{Fl\'avia B. Ramos}
	\affiliation{Physics Department and Research Center OPTIMAS, University of Kaiserslautern-Landau, 67663 Kaiserslautern, Germany}
	\author{Andrew Urichuk}
\affiliation{Department of Physics and Astronomy and Manitoba Quantum Institute, University of Manitoba, Winnipeg R3T 2N2, Canada}
	\affiliation{Fakult\"at f\"ur Mathematik und Naturwissenschaften,
		Bergische Universit\"at Wuppertal, 42097 Wuppertal, Germany}
		\author{Imke Schneider}
  \affiliation{Physics Department and Research Center OPTIMAS, Rheinland Pf\"alzische Technische Universit\"at, 67663 Kaiserslautern, Germany}
	\author{Jesko Sirker}
	\affiliation{Department of Physics and Astronomy and Manitoba Quantum Institute, University of Manitoba, Winnipeg R3T 2N2, Canada}

	\begin{abstract}
	 We investigate the relaxation dynamics of equal-time correlations 
	 in the antiferromagnetic phase of the XXZ spin-1/2 chain 
	 following a global quantum quench of the anisotropy parameter. We focus, in particular, on the relaxation dynamics starting from an initial N\'eel state. Using state-of-the-art density-matrix renormalization group simulations, the exact solution of an effective free-fermion model, and the quench-action approach within the thermodynamic Bethe ansatz, we show that the late-time relaxation is characterized by a power-law decay $\sim t^{-3/2}$ independent of anisotropy. This is in contrast to the previously studied exponential decay of the antiferromagnetic order parameter.
	 Remarkably, the effective model describes the numerical data extremely well even on a quantitative level if higher order corrections to the leading asymptotic behavior are taken into account.
	\end{abstract}

\date\today

\maketitle

	\section{Introduction}
	The high control available in experiments on ultracold atomic gases has opened the door to investigate non-equilibrium dynamics in many fundamental quantum models of Statistical Physics \cite{Kinoshita2004,Paredes2004,Bloch2005,Kinoshita2006,Sadler2006,Hofferberth2007,RevModPhys.82.1225,Trotzky2012,Cheneau2012,jepsen_spin_2020}. Probing relaxation and thermalization becomes possible in these systems because decoherence and dissipation remain small for sufficiently long time scales. The typical out-of-equilibrium scenario in cold atomic gases is the so-called quantum quench where a control parameter is suddenly changed. 
	
	From a theoretical  point of view, investigating how quantum many-body systems equilibrate is an enormous challenge. Except for Gaussian problems, which can either be solved analytically or at least simulated numerically for long times, the exponential growth of the Hilbert space with the system size limits exact diagonalizations to small systems. Since recurrence times are relatively short in small systems, it is often very difficult to draw any conclusions from such calculations about the thermodynamic limit. On the other hand, methods based on a truncation of the Hilbert space such as the density-matrix renormalization group (DMRG) \cite{PhysRevLett.93.076401,Enss2012} or, more generally speaking, approaches based on matrix-product states or operators, suffer from accuracy loss due to the growth of the entanglement entropy (EE) with time. For global quenches in clean one-dimensional systems, in particular, the EE grows linearly in time which can be understood in terms of quasiparticles spreading in a light-cone-like fashion \cite{Calabrese2005}.
	
	The stationary state of a system is determined by its (quasi-)local conserved charges. If the system is non-integrable then its subsystems are generally expected to relax to a density matrix described by a Gibbs ensemble with an effective temperature set by the energy density \cite{PhysRevA.43.2046,PhysRevE.50.888,Rigol2008,SirkerKonstantinidis}. In contrast, the infinite number of local conserved charges in an integrable model prevent thermalization. In this case, a relaxation to stationary values given by generalized Gibbs ensembles (GGE) is expected \cite{PhysRevLett.98.050405,PhysRevA.74.053616,Langen2015}. Quench dynamics has been studied in the context of effective free models, Luttinger liquids, conformal field theories, and interacting systems, see for example Refs.~\cite{PhysRevLett.100.100601,PhysRevB.79.155104,PhysRevA.80.063619,PhysRevLett.97.156403,PhysRevLett.102.130603,Barmettler_2010,Cramer2010,PhysRevLett.106.140405,PhysRevLett.106.227203,Calabrese20121,Calabrese20122,PhysRevLett.110.257203,Brockmann2014,Pereira_2014,Bucciantini_2014,Karrasch_2015,PhysRevB.89.165104,PhysRevB.94.075129,Calabrese2016,10.21468/SciPostPhysLectNotes.17,PhysRevB.101.041110,PhysRevLett.126.210602,10.21468/SciPostPhys.11.6.104,10.21468/SciPostPhys.11.3.055,PhysRevB.105.195103,10.21468/SciPostPhys.12.5.158}.
	
	While the relaxation dynamics after a quench is by now quite well understood in Gaussian models, much less is known for interacting systems. The XXZ chain, in particular, is well suited to serve as a simple model for the investigation of interaction effects. Due to its integrability, methods based on the Bethe ansatz can be used \cite{PhysRevLett.110.257203,PhysRevLett.113.117203,Brockmann_2014,Caux_2016,Alba_2016,PhysRevLett.113.117203}. In this context, one important method is the quench-action approach (QA): If the form factors of the initial state with the Bethe eigenstates are known, this method can be used to calculate the stationary values of local observables at long times after the quench. So far, however, the relaxation dynamics leading to equilibration has remained out of reach of Bethe ansatz based approaches.
	
	In this work, we investigate the time-dependent behavior of equal-time correlators in the XXZ chain after an interaction quench within the gapped antiferromagnetic phase. More specifically, we consider the quench from the classical N\'eel state (the ground state of the XXZ chain for infinite anisotropy) to a finite anisotropy. Using the state-of-art light-cone renormalization group (LCRG) \cite{Enss2012}, we obtain numerical results in which the long-time regime of local correlation functions can be studied. On the analytical side, inspired by Barmettler \emph{et al.} \cite{PhysRevLett.102.130603,Barmettler_2010}, we consider the large-anisotropy limit, in which the time evolution of the system can be related to the dynamics governed by the XZ chain. In this effective framework, we obtain exact formulas for the long-time behavior of short-distance correlators. In contrast to the non-oscillatory exponential decay of the order parameter \cite{PhysRevLett.102.130603,Barmettler_2010},	we show based on a saddle-point analysis that the correlations decay asymptotically in a power-law fashion with  interaction-independent exponents. In addition, we use the QA method to compute the stationary values of the time-dependent correlations in the XXZ model. Finally, we define an asymptotic formula, based on the analytical results for the effective model, to fit the LCRG data. A remarkable, quantitative agreement between the XZ results with renormalized parameters and the numerical results for the XXZ chain is reported.
	
	This paper is organized as follows: In Sec.~\ref{model}, we introduce the model, quench protocol, and methods we use to investigate the problem of interest. We devote Sec.~\ref{strongd} to the discussion of the large-anisotropy limit. Here, we map the interacting problem to a spinless free-fermion model and derive its exact solution as well as the asymptotic behavior of short-distance correlators.  In Sec.~\ref{numerics}, we present our LCRG results for three different correlators and discuss them in terms of the effective theory. Finally, we provide concluding remarks in Sec.~\ref{conclusion}.

	\section{Model and methods}
	\label{model}
	We consider the isolated XXZ chain described by the Hamiltonian
	\be
	H(\Delta)=J\sum_{j}\left( \sigma^x_{j}\sigma^x_{j+1}+\sigma^y_{j}\sigma^y_{j+1}+\Delta \sigma_j^z\sigma^z_{j+1}\right),\label{hxxz}
	\ee
	where $\sigma^{x,y,z}_j$ are the Pauli matrices acting at site $j$. $J>0$ is the exchange coupling and $\Delta$ is the anisotropy parameter. This model exhibits a U(1) symmetry corresponding to  invariance under rotation by an arbitrary angle around the $z-$axis. Furthermore, the XXZ chain is exactly solvable by Bethe ansatz and its ground-state phase diagram is characterized by a gapless  phase with quasi-long-range order for $|\Delta |\le 1$  and a gapped antiferromagnetic (ferromagnetic) phase for $\Delta>1$ $(\Delta<-1)$ \cite{Bethe1934}. The limits $\Delta\rightarrow\infty$ and  $\Delta\rightarrow-\infty$ correspond to the classical Ising  antiferromagnet and ferromagnet, respectively. 
	
	Here, we will focus on the relaxation dynamics after a global quantum quench of the anisotropy in the antiferromagnetic phase of the XXZ chain. The quench protocol is as follows: The system is prepared in the N\'eel state defined as
	\be
	|\psi_N\ket=|\uparrow\downarrow\uparrow\downarrow\cdots\uparrow\downarrow\rangle,\label{neel}
	\ee
	which is a ground state of $H(\infty)$. At time $t=0$, the anisotropy parameter is suddenly quenched to a finite  $\Delta$, so that $|\psi_N\rangle$ is no longer an eigenstate of $H(\Delta)$. The unitary time evolution of the system is governed by the time-dependent Schr\"odinger equation
	\be
	|\psi(t)\rangle=e^{-iH(\Delta)t}|\psi_N\rangle.\label{timeevol}
	\ee
	We focus, in particular, on the equal-time correlations along the $z$-direction at time $t$ after the quench which are given by
	\begin{equation}
	C(\ell,t)=\bra \psi_N|\sigma_0^z(t) \sigma_\ell^z(t)|\psi_N\ket.\label{correl}
	\end{equation}
	
	The number of excitations (spin flips) on top of the N\'eel state depends on the anisotropy parameter $\Delta$ during the time evolution. For $\Delta\gg 1$,  the N\'eel state is close to the ground state of $H(\Delta)$ so that the density of excitations during the time evolution is expected to be low. If only few spin flips (kinks) are present, the time evolution can be approximated by that of an effective XZ chain \cite{LIEB1961407,PhysRevA.3.786,PhysRevA.69.053616}. This effective model can be mapped onto a Gaussian fermionic model and the long-time behavior of $C(\ell,t)$ can be extracted from a saddle-point analysis (see Sec.~\ref{strongd}). In addition, the stationary values for the full XXZ chain \eqref{hxxz} can be calculated using the QA approach. Some stationary values of short distance correlators at large $\Delta$ are already available from Ref.~\cite{Brockmann_2014}.

	We will numerically calculate the time dependence of local operators in the thermodynamic limit, e.g. $C(\ell,t)$, using the LCRG method. In contrast to the infinite time-evolving block decimation, LCRG does not require translational invariance. The method relies on a Suzuki-Trotter (ST) decomposition of the time evolution operator and a systematic truncation of the Hilbert space  that is iteratively carried out using a reduced density-matrix \cite{Enss2012}. By keeping the effective velocity due to the ST decomposition much larger than the velocity of excitations, the Lieb-Robinson bound guarantees that the system is effectively always in the thermodynamic limit. In comparison to the time-dependent DMRG, the major advantage of LCRG lies in the computational speed-up achieved due to the light-cone structure of transfer matrices, which allows for a reduction of bond operators in the ST decomposition.
	
	As already mentioned in the introduction, the entanglement imposes limitations on the simulation times achievable by DMRG-based methods.
	Keeping up to 65,000 states to represent the truncated Hilbert space and using a second-order ST decomposition, we can reach times $Jt\lesssim5$ while keeping the truncation error below $10^{-9}$. It is worth keeping in mind that if one wants to compare this simulation time with previous studies that we are using Pauli matrices in the definition of the Hamiltonian instead of spin-$1/2$ operators. If we rewrite the Hamiltonian \eqref{hxxz} in terms of spin operators, then our simulation times correspond to $Jt\lesssim 20$, i.e. in this case $t\rightarrow 4t$. The numerical results are more sensitive to the finite Trotter step $\delta t$ the deeper one goes into the massive phase. This is because the amplitude of oscillations in the correlators becomes smaller with increasing $\Delta$ while the oscillation frequency increases. By analyzing the convergence of our numerical results for different values of $\delta t$, we find that accurate results are obtained for $\delta t=0.005$ $(\delta t=0.001)$ if $\Delta\leq 12$  $(\Delta>12)$. The Trotter error for the decomposition used is always of order $(\delta t)^2$. Additional numerical details are reported in Sec.~\ref{numerics}.
	
	Let us now briefly comment on how stationary values are obtained from the QA approach. The overlap between the initial state, here the N\'eel state, and Bethe states can be determined ~\cite{Brockmann2014,PhysRevLett.113.117203} and inserted into the quench action. Once these overlaps are known, the minimization of the quench action yields integral relations---known as the overlap thermodynamic Bethe ansatz (TBA) equations---which determine the saddle-point string densities~\cite{Brockmann_2014,mestyan_2015} as
	\be
0 = \frac{\delta S_{\text{QA}}[\vec \rho]}{\delta \rho_n} \bigg|_{\vec \rho = \vec \rho^{\text{sp}}} \quad \text{for} \quad n \geq 1.
	\ee
	These saddle-point solutions are combined with quantum transfer matrix (QTM) formulas to determine the long-time limit as 
	\be
	\lim_{t \to \infty} C(\ell,t) = \langle \rho^{\text{sp}} | \sigma_0^z \sigma_\ell^z | \rho^{\text{sp}} \rangle.
	\ee
More details on the quench-action approach are discussed in App.~\ref{QAA}. In order to have a guide to analyze our numerical results, we start with the analytical investigation of the strong-anisotropy limit.
	
	\section{Strong-anisotropy limit} \label{strongd}
	Using the raising and lowering operators $\sigma^{\pm} _j= (\sigma_j^x \pm i\sigma_j^y)/2$ we can write $\sigma^x_j\sigma^x_{j+1}+\sigma^y_j\sigma^y_{j+1}=2(\sigma^+_j\sigma^-_{j+1}+\sigma^-_j\sigma^+_{j+1})$ while $2\sigma^x_j\sigma^x_{j+1}=2(\sigma^+_j\sigma^-_{j+1}+\sigma^-_j\sigma^+_{j+1})+2(\sigma^+_j\sigma^+_{j+1}+\sigma^-_j\sigma^-_{j+1})$. For $\Delta\gg 1$, the quench will introduce a density of kinks ($\pi$ phase shifts between N\'eel ordered regions) which remains small during the entire time evolution so that replacing $\sigma^x_j\sigma^x_{j+1}+\sigma^y_j\sigma^y_{j+1}$ by $2\sigma^x_j\sigma^x_{j+1}$ is a reasonable approximation \footnote{Note that we could as well replace $\sigma^x_j\sigma^x_{j+1}+\sigma^y_j\sigma^y_{j+1}$ by $2\sigma^y_j\sigma^y_{j+1}$ which leads to the same effective Hamiltonian after a rotation around the $z$-axis which is a symmetry of the Hamiltonian.}. Thereby, the XXZ chain effectively becomes the XZ model, whose Hamiltonian is
	\be
	H_{{XZ}}=J\sum_{j}\left( 2\sigma^x_j\sigma^x_{j+1}+\Delta \sigma^z_j\sigma^z_{j+1}\right).\label{hxz}
	\ee 
	We note that this approximation leads to an explicit U(1)-symmetry breaking due to the addition of the terms $\sigma^+_j\sigma^+_{j+1}$ and  $\sigma^-_j\sigma^-_{j+1}$ which do not conserve the total magnetization. 
	
	The XZ chain can be mapped to the exactly solvable XY chain by a  90$^\circ$-degree rotation around the $x$-axis. After the rotation, the Hamiltonian in Eq. (\ref{hxz}) can be expressed in terms of the ladder operators as
	\begin{eqnarray}
H_{{XZ}}=	J\sum_{j}\left[(2+\Delta)(\sigma^+_j\sigma^-_{j+1}+\sigma^-_j\sigma^+_{j+1})\right.\nonumber\\
	\left.+(2-\Delta)(\sigma^+_j\sigma^+_{j+1}+\sigma^-_j\sigma^-_{j+1})\right] .\label{hxzpm}
	\end{eqnarray}
	The system described by Eq.~\eqref{hxzpm} corresponds to a spinless free-fermion model and its exact solution is obtained by means of the Jordan-Wigner transformation
	\be
	\sigma^+_j\rightarrow c^\dagger_je^{i\pi \phi_j },\quad \sigma^-_j\rightarrow c_j e^{-i\pi \phi_j },\label{jwtransf}
	\ee
	where $c^\dagger_j$ and $c_j$ are the fermionic creation and annihilation operators acting at site $j$ and  $\phi_j=\sum_{\ell=0}^{j-1}c^\dagger_\ell c_\ell $ is the Jordan-Wigner string.
	In Fourier space, the fermionic Hamiltonian can then be diagonalized by performing a Bogoliubov transformation that maps the operators $c_k=\frac{1}{\sqrt{L}}\sum_j e^{-i\pi j}  c_j$, where $L$ is the system size, to a new set of fermionic operators $a_k$. The corresponding transformation $U_\Delta$ is defined by
	\begin{align}
	\left(\begin{array}{c}
	c_{-k}\\
	c_{k}^{\dagger}
	\end{array}\right)=U_\Delta\left(\begin{array}{c}
	a_{-k}\\
	a_{k}^{\dagger}
	\end{array}\right)\nonumber,\\*[0.1cm]
	U_\Delta=\left(\begin{array}{cc}
	\cos\theta_k & -i\sin\theta_k\\
	-i\sin\theta_k & \cos\theta_k
	\end{array}\right),
	\end{align}
	with $\tan{\left(2\theta_k\right)=\frac{2-\Delta}{2+\Delta}}\tan (k)$. The Hamiltonian in the new operators then reads
	\be
	H_{XZ}=\frac{1}{2}\sum_{k}\left(\begin{array}{cc}
		a_{k}^{\dagger} & a_{-k}\end{array}\right)\left(\begin{array}{cc}
		\varepsilon_{k} & 0\\
		0 & -\varepsilon_{k}
	\end{array}\right)\left(\begin{array}{c}
		a_{k}\\
		a_{-k}^{\dagger}
	\end{array}\right),
	\ee
	where $\varepsilon_k=4J\sqrt{1+\Delta^2/4+\Delta \cos(2k)}$ is the dispersion relation. Hence, the ground-state phase diagram of the system is characterized by a critical point at $\Delta=2$ separating two antiferromagnetic phases with alignment along the $x-$ and $z-$direction for $0<\Delta<2$ and $\Delta\ge 2$, respectively. Note that the classical N\'eel state, $|\psi_N\ket$, is the ground state of $H_{XZ}$ for $\Delta=\infty$.
	
	Let us now turn our attention to the asymptotic behavior of $C(\ell,t)$. Using the transformation in Eq.~(\ref{jwtransf}), $C(\ell,t)$ is written as
	\be
	C(\ell,t)=-\bra ( c^\dagger_0 - c_0)(c^\dagger_\ell e^{i\pi\phi_\ell}-e^{-i\pi \phi_\ell}c_\ell) \ket.\label{corrjw}
	\ee
	Defining $A_j=(c^\dagger_j+c_j)$ and $B_j=(c^\dagger_j-c_j)$, and observing that $e^{i\pi c^\dagger_jc_j}=-B_jA_j$, we have
	\be
	C(\ell,t)=(-1)^\ell\bra A_0B_1A_1\cdots A_{\ell-1}B_\ell \ket.\label{corrma}
	\ee
	
	The $\ell$-point correlation can then be decomposed into a combination of pairwise contractions via Wick's theorem \cite{Caianiello1958}. Thereby, the problem is reduced to calculating the contractions $\bra A_mA_n \ket$, $\bra A_mB_n \ket$, $\bra B_mA_n \ket$, and $\bra B_mB_n \ket$. To do so, it is convenient to write the time-dependent fermionic operators in the basis which diagonalizes $H_{XZ}$ for $\Delta=\infty$
	\be
	\left(\begin{array}{c}
		c_{-k}(t)\\
		c_{k}^{\dagger}(t)
	\end{array}\right)=U_{\Delta}\left(\begin{array}{cc}
		e^{-i\varepsilon_{k}t} & 0\\
		0 & e^{i\varepsilon_{k}t}
	\end{array}\right)U_{\Delta}^{-1}U_{\infty}\left(\begin{array}{c}
		a_{-k}^{\infty}\\
		a_{k}^{\infty\dagger}
	\end{array}\right),
	\ee
	so that  $a^\infty_k|\psi_N\ket=0$. For $m\neq n$, the aforementioned contractions can then be written as 
	\bea
	\bra A_m A_n \ket &=&\int_{-\pi}^{\pi}\frac{dk}{2\pi}e^{-ik(m-n)}\left[1-\sin{(2\varepsilon_kt)}\sin{(2\tilde{\theta}_k)}\right], \nonumber \\
	\bra A_m B_n \ket &=& \int_{-\pi}^{\pi}\frac{dk}{2\pi}e^{-i[k(m-n)+2\theta_k]}\left[\right.\cos(2\tilde\theta_k)\nonumber\\
	&&\qquad +i\cos{(2\varepsilon_kt)\sin(2\tilde\theta_k)}\left.\right],\label{ambn}
	\eea     
	where $\tilde\theta_k=\theta_k-\theta_k^\infty$ with $\theta_k^\infty=-k/2$. Besides, we have $\langle B_mB_n \rangle = \langle A_mA_n\rangle $ and $\langle A_mB_n\rangle=-\langle B_nA_m\rangle$.
	More details about the calculations discussed in this section can be found in App.~\ref{asymapp}. For simplicity, we will set $J=1$ as the energy scale henceforth.
	
	Now, we focus on the distance-1 correlator, 
		\bea
	C(1,t)=-\frac{1}{2\pi}\int_{-\pi}^{\pi}dk \, e^{i(2\theta_k+k)}   \left[\cos(2\theta_k+k)\right.\nonumber\\
	-i\left.\cos(2\varepsilon_kt)\sin(2\theta_k+k)\right].\label{c1t}
	\eea

	We use the saddle-point method to determine the asymptotics of $C(1,t)$. By deforming the complex contour in the vicinity of the roots of $\varepsilon_k'=0$, this method provides the leading contributions to the long-time regime. In particular, by expanding $\varepsilon_k$ up to fourth order around the inflection points, we obtain
	\begin{widetext}
		\begin{eqnarray}
	C(1,t)&\approx&-\left(1-\frac{2}{\Delta^2}\right)-\frac{\left(1+\frac{2}{\Delta}\right)^{1/2}}{32\sqrt\pi(1+\Delta^2/4+\Delta)}\left\{\left(1+\frac{2}{\Delta}\right)\frac{\sin\left[\tilde\omega_1(\Delta)t-\frac{\pi}{4} \right]}{t^{3/2}}+\frac{3(1+\Delta^2/4-\Delta/2)}{16\Delta^2}\right.\nonumber\\
	&\times &\left.\frac{\cos\left[\tilde\omega_1(\Delta)t-\frac{\pi}{4}\right]}{t^{5/2}}\right\}-\frac{\left(1-\frac{2}{\Delta}\right)^{1/2}}{32\sqrt\pi(1+\Delta^2/4-\Delta)}\left\{\left(1-\frac{2}{\Delta}\right)\frac{\sin\left[\tilde\omega_2(\Delta)t-\frac{3\pi}{4} \right]}{t^{3/2}}\right.\nonumber\\ &+&\frac{3(1+\Delta^2/4+\Delta/2)}{16\Delta^2}\left.\frac{\cos\left[\tilde\omega_2(\Delta)t-\frac{3\pi}{4}\right]}{t^{5/2}}\right\}+O(t^{-7/2}),\label{asym1}
	\end{eqnarray}
	\end{widetext}
where $\tilde\omega_1(\Delta)=8(\Delta/2+1)$ and $\tilde\omega_2(\Delta)=8(\Delta/2-1)$.
	The first term in Eq.~\eqref{asym1} is the stationary value of $C(1,t)$. For $\Delta\rightarrow\infty$, this result is consistent with the energy density of the classical N\'eel state. Moreover, we observe that the time-dependent part contains two oscillation frequencies proportional to the anisotropy and two exponents, independent of anisotropy, that define the dominant power-law decay of the correlation function.
	
	Following the same rationale, one can obtain distance $\ell=2$  and $\ell=3$ correlators. It turns out that $C(2,t)$ is particularly simple because only $\bra A_0B_1\ket\bra A_1 B_2 \ket$ does not vanish after applying Wick's theorem. In this case, $C(2,t)=\left(C(1,t)\right)^2$. On the other hand, for $C(3,t)$, two non-vanishing terms remain
	\be
	C(3,t)=-\bra A_0B_1\ket^3+\bra A_0B_1\ket\bra A_2B_1 \ket\bra A_0B_3 \ket.	\label{c3t}\ee  
	Up to fourth order in the saddle-point analysis, each contraction in Eq.~\eqref{c3t} contributes with one time-independent term, two terms with distinct frequencies that decay with power-law exponent $3/2$ and two other with exponent $5/2$ (see App. \ref{asymapp}). This implies that the correlation function $C(\ell,t)$ always decays to leading order as $1/t^{3/2}$ in the asymptotic long-time limit independent of the distance $\ell$. For the distance three case we find, in particular, that the saturation value $C(3,\infty)$ is given by
	\begin{equation}
	C(3,\infty)\approx -\left(1-\frac{2}{\Delta^2}\right)^3-\frac{\left(\Delta^2 -2 \right)(\Delta^2 - 4)}{\Delta^6}.\label{c3inf}
	\end{equation} 
	
	It is worth mentioning that in contrast to short-distance correlators, the relaxation of the order parameter is difficult to calculate because it is nonlocal in the fermionic basis. However,	a numerical evaluation for large system sizes is possible \cite{PhysRevLett.102.130603,Barmettler_2010} and Barmettler \emph{et al.} find that the long-time behavior of the staggered magnetization for $\Delta \geq 2$ shows a non-oscillatory exponential decay with a relaxation time scale proportional to $\Delta^2$. It is also worth noting that the time-dependent behavior of correlators for a quench within the ferromagnetic phase of the transverse-field Ising chain was studied in Ref.~\cite{Calabrese20122}.
	
	\section{Numerical results}\label{numerics}
	\begin{figure}
		\centering{\includegraphics[width=8cm]{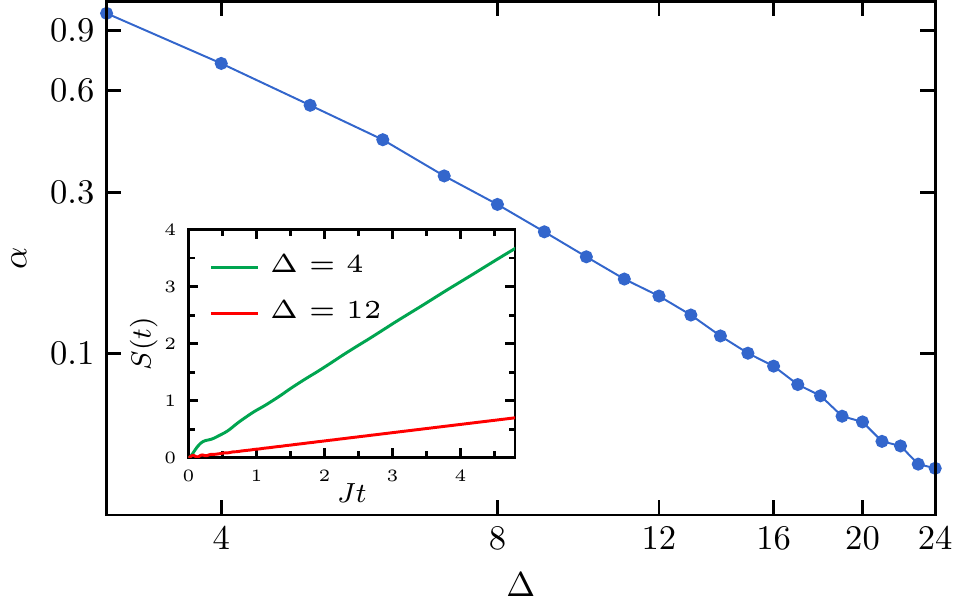}}
		\caption{Log-log scale plot of the slope $\alpha$ extracted from the linear growth of the EE as a function of $\Delta$. The inset shows typical results of the time dependence of $S(t)$ for $\Delta=4$ and $12$.}
		\label{entt}
	\end{figure}
\begin{figure}
	\centering{\includegraphics[width=8cm]{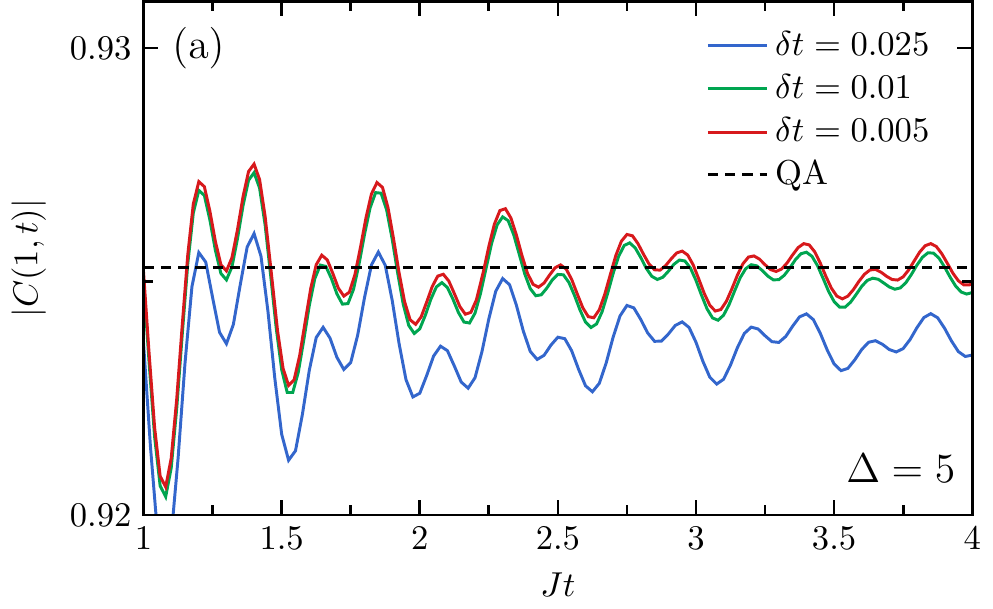}}
	\centering{\includegraphics[width=8cm]{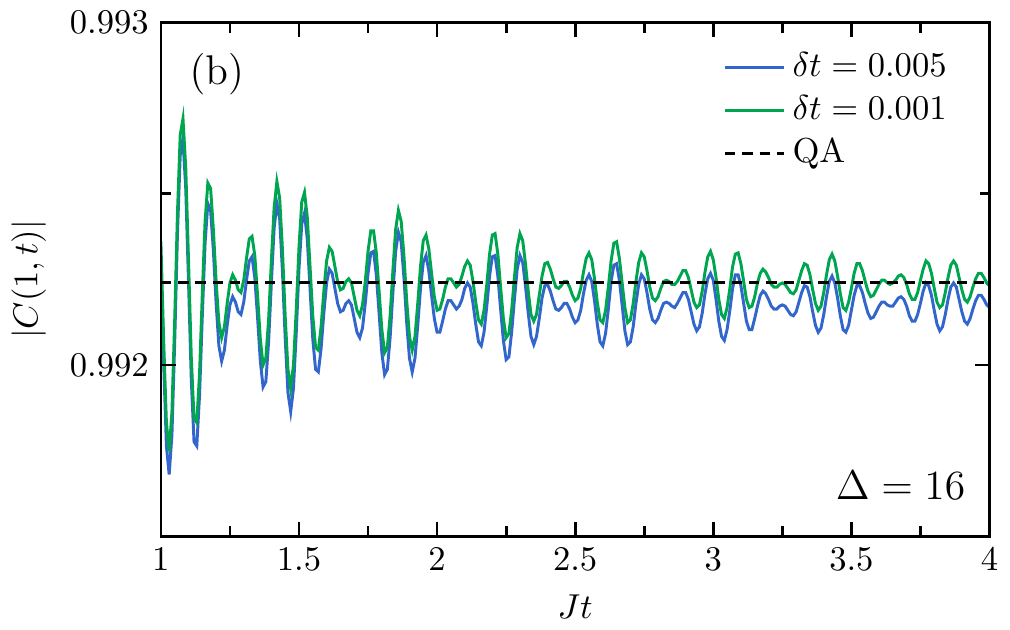}}
	\caption{Correlator $C(1,t)$ as a function of time for different Trotter steps and anisotropies: (a) $\Delta=5$ and (b) $\Delta=16$. The dashed lines indicate the stationary values obtained from the QA approach.}
	\label{ddt}
\end{figure}
	In this section, we present numerical results for the quench dynamics of the XXZ chain obtained by the LCRG. As discussed in the previous section, in the massive regime the quench dynamics is approximately governed by the effective XZ model for which the time-dependent correlations can be exactly calculated via a fermionization. Furthermore, the saturation values of  $C(\ell,t)$ can be directly computed for the XXZ model using the QA method. We are therefore in a position where we can provide a systematic analysis of the LCRG data for the quench dynamics.
	
	We start discussing the two main sources of errors in our numerical results: the truncation error and the Trotterization of the time evolution operator. To keep the largest truncation error of order $10^{-9}$ or below at late times, we include up to $65,000$ states in the truncated Hilbert space. 
	We observe that longer simulation times can be reached for larger anisotropy $\Delta$. This trend can be understood by considering the bipartite entanglement entropy $S(t)$ which grows linearly in time, $S=\alpha t$. As can be seen in Fig.~\ref{entt}, the slope $\alpha$ is decreasing approximately in a power-law fashion. The typical time dependence of $S(t)$ is shown in the inset of Fig.~\ref{entt} for two distinct values of $\Delta$. On the other hand, the oscillation frequency increases with $\Delta$ while the amplitude of the oscillation decreases. Thus, the time step (Trotterization) plays an important role for the accuracy of the simulations at strong anisotropy. In Fig.~\ref{ddt}, we show how our numerical results depend on $\delta t$. Stationary values obtained from the QA are also indicated in the figure. We use these exact values to determine a time step $\delta t$ which leads to errors that are small compared to the amplitude of the oscillations.
\begin{figure}
\centering{\includegraphics[width=8cm]{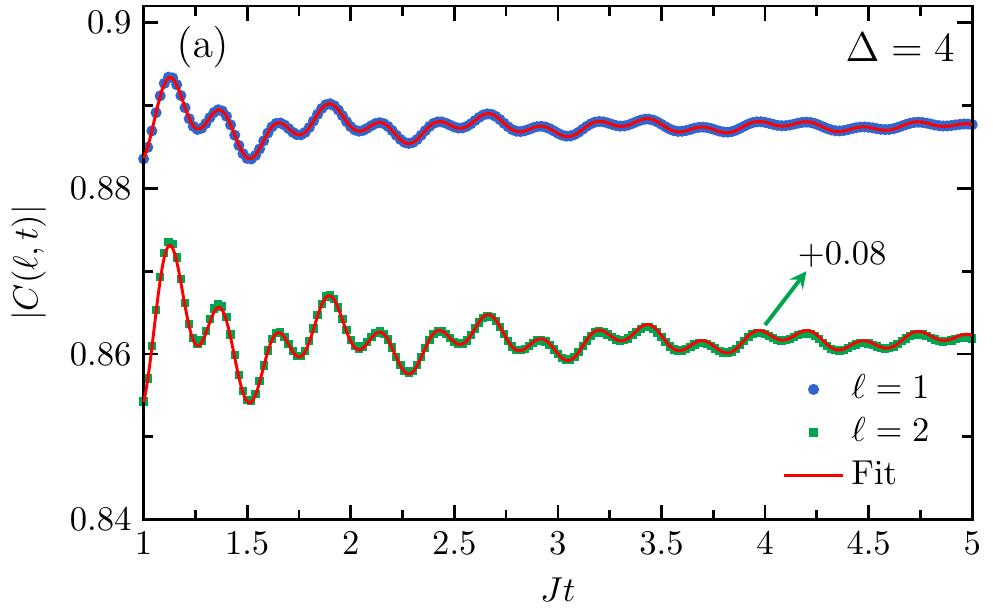}}
\centering{	\includegraphics[width=8cm]{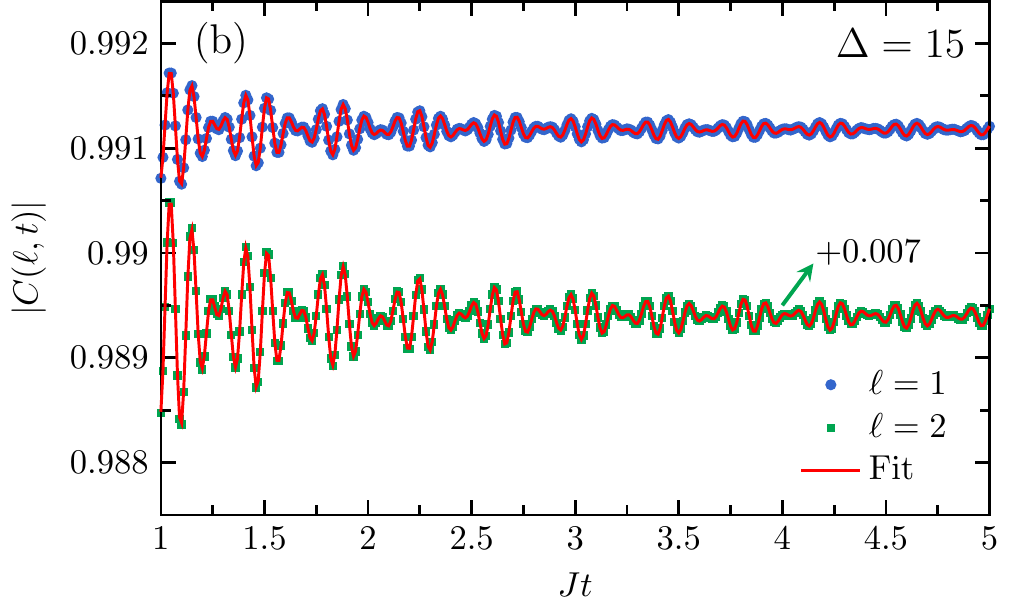}}
\centering{	\includegraphics[width=8cm]{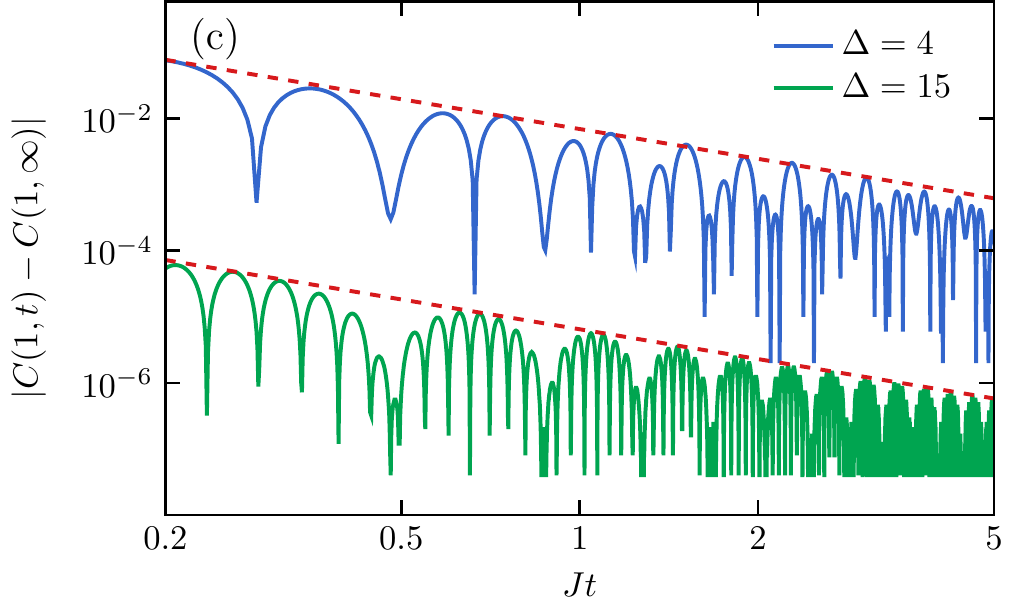}}
\caption{Equal-time correlations for distances $\ell=1,2$ and (a) $\Delta=4$, (b) $\Delta=15$. The symbols are LCRG results and the red solid lines are fits to our data using Eq. (\ref{fitmass}). To display the results for both distances in the same plot, we added an offset to $C(2,t)$ as indicated. (c) Oscillatory part of the distance-1 correlator on a log-log scale for $\Delta=4$ and 15. The dashed red lines are power laws with exponent $-3/2$.}
\label{fig1}
\end{figure}
\begin{figure}
	\centering{}
	\includegraphics[width=8cm]{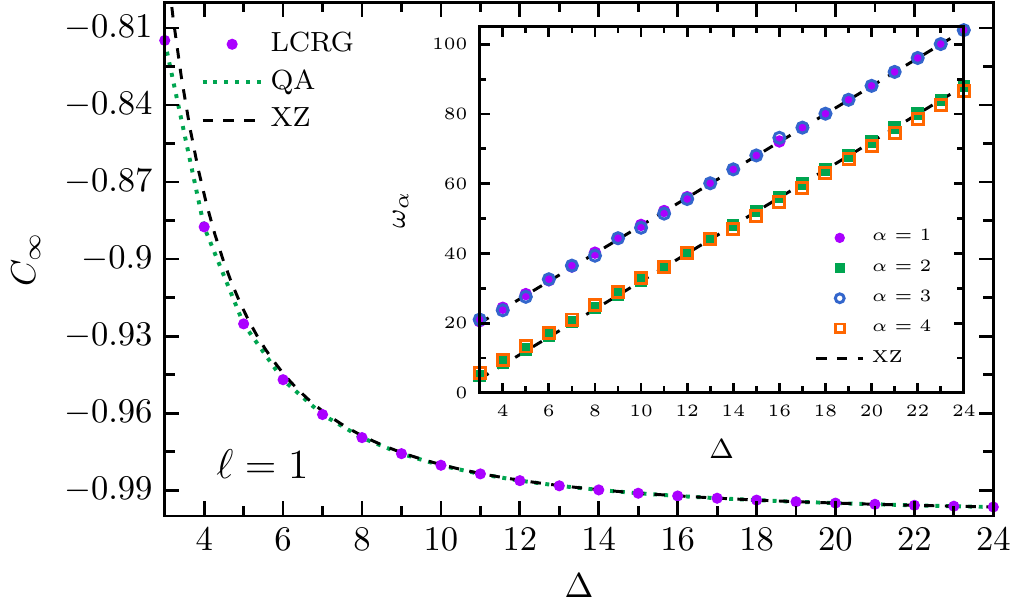}
	\caption{Stationary values of $C(1,t)$ obtained by fitting the LCRG data, using QA, and the exact solution of the XZ chain. The inset shows estimates of the oscillation frequencies $\omega_\alpha$ as a function of $\Delta$ for $C(1,t)$. The analytical results $\omega_{1,3}=8(\Delta/2+1)$ and  $\omega_{2,4}=8(\Delta/2-1)$ for the strong-anisotropy regime are also shown in the plot.} \label{fig2}
\end{figure}	

Let us now discuss the temporal behavior of $C(\ell,t)$. In Fig.~\ref{fig1}, we show the time-dependent correlations for $\Delta=4$ and $\Delta=15$. Using the analytical results for $C_\infty=\lim_{t\to\infty} C(\ell,t)$ from the QA and the asymptotics for the XZ model discussed in Sec.~\ref{strongd}, we define a function to fit distance-1 and distance-2 correlators
\begin{eqnarray}
|C(\ell,t)|&=&\left\{ C_{\infty} +  \frac{A_1\sin(\omega_1 t+\phi_1)+A_2\sin(\omega_2 t+\phi_2) }{t^{3/2}}\right.\nonumber\\
&+&\left.  \frac{A_3\cos(\omega_3 t+\phi_3)+A_4\cos(\omega_4 t+\phi_4) }{t^{5/2}}\right\}^\ell. \label{fitmass}
\end{eqnarray}
We keep the exponents of the leading and subleading terms fixed to reduce the number of free parameters in the fitting equation and because we do not expect them to change as compared to the XZ model. We notice that close estimates of these exponents are obtained if they are kept as free parameters.  Note that Eq.~(\ref{fitmass}) perfectly fits the time dependence of the LCRG results. Similar high-quality fits can be obtained for all $\Delta\gtrsim 3$. A clear power-law decay is observed in Fig.~\ref{fig1}(c). The red dashed lines in the figure indicate that the leading time-dependent contribution decays with exponent 3/2. In Fig.~\ref{fig2}, we summarize the estimates for $\omega_{1-4}$  and $C_\infty$ obtained from such fits as a function of $\Delta$ for the distance-1 correlator. A comparison with the exact solution of the XZ chain and the QA values for $C_\infty$ is also shown. Overall, we observe an excellent agreement between the fitted oscillation frequencies for the XXZ model and those for the exactly solvable XZ chain. The fitted stationary values $C_\infty$ are slightly shifted as compared to those for the XZ model but agree very well with the exact QA results. 
Very similar results are obtained for the case of $\ell=2$. It is worth stressing that the remarkable agreement for all shown $\Delta$ values is surprising given that the effective theory might only be expected to hold at very large $\Delta$. To summarize, we observe an excellent agreement with the predictions of the XZ model with a small, $\Delta$-dependent shift in the stationary value---fully covered and understood by the QA approach---as well as a renormalization of the oscillation amplitudes. The oscillation frequencies, on the other hand, remain essentially unchanged.

The analysis of the regime $1<\Delta<3$ is challenging. First, in the regime $\Delta>2$, where the XZ model remains ordered along the $z-$direction, higher order corrections to the asymptotics derived in Sec.~\ref{strongd} become more important and longer times are needed before they can be neglected. This can be seen by comparing the asympotics with a numerical integration of Eq.~\eqref{c1t}. Reaching much longer time scales is, however, not possible with current algorithms and computing resources. For $1<\Delta<2$, there might also be additional contributions to the asymptotics which are not captured by the XZ model.

\begin{figure}
	\includegraphics[width=8cm]{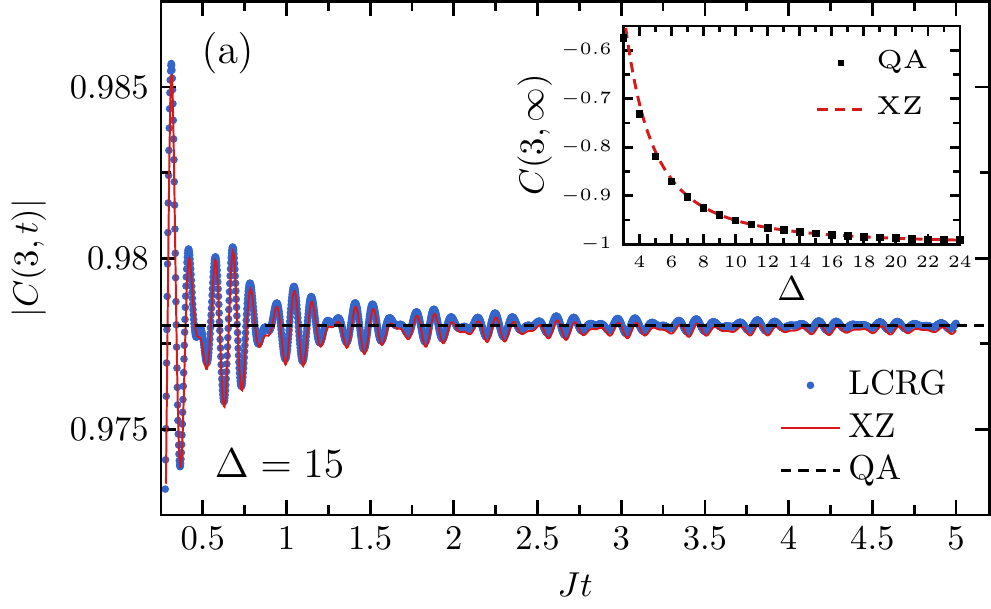}
	\includegraphics[width=8cm]{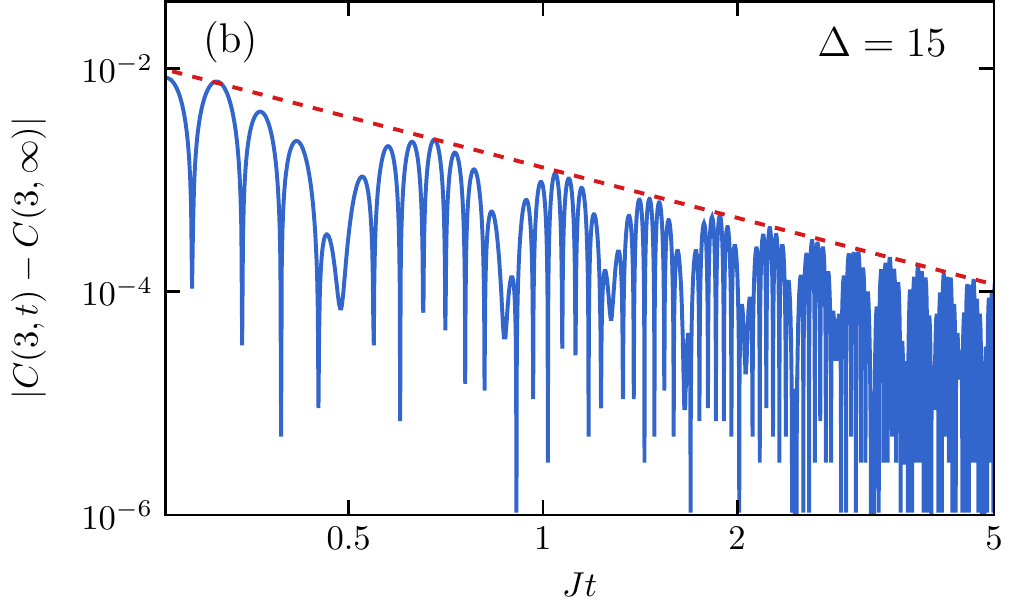} 
\caption{(a) Equal-time correlation $C(3,t)$ for $\Delta=15$. The filled circles are LCRG results and the solid red line is the result for the XZ model, Eq.~\eqref{c3t} [see also Eqs.~\eqref{as0}-\eqref{fullcont} in App.~\ref{asymapp}]. The black dashed line is the exact stationary value obtained from QA. The inset compares the QA stationary values for the XXZ chain and the stationary value for the XZ chain, Eq.~\eqref{c3inf}. (b) Corresponding oscillatory part: The red dashed line is a power law with an exponent $-3/2$.}

	\label{fig3}
\end{figure}

	Finally, let us also discuss the distance-3 correlator. In this case, due to the combination of different Wick contractions, many free parameters would be required to define a fit function and achieving convergence of the fit becomes more difficult. Instead, we concentrate on large $\Delta$ where we can directly evaluate Eq.~\eqref{c3t} by a numerical integration. In Fig.~\ref{fig3}(a), we show a typical result for such a correlator for $\Delta=15$ and compare it directly to the XZ model and the stationary value obtained by the QA. In the inset of Fig.~\ref{fig3}(a), we show that the stationary value obtained for the XZ model, Eq.~\eqref{c3inf}, agrees well with the QA result. In addition, in Fig.\ref{fig3}(b), the oscillatory part of $C(3,t)$ is shown on a log-log scale supporting the XZ-chain result of a power-law decay with exponent $-3/2$.
 
	\section{Conclusions} \label{conclusion}
	We have studied the relaxation dynamics after a global quantum quench in the massive antiferromagnetic phase of the XXZ chain using the N\'eel state as initial state. By adding terms that explicitly violate the conservation of the total magnetization in the XXZ model, we can obtain a XZ model which is exactly solvable via a Jordan-Wigner transformation. We have argued that at large anisotropy $\Delta$, the terms violating the conservation of the total magnetization have little effect on the dynamics. The XZ model is thus a useful effective model for which the long-time asymptotics of correlation functions can be obtained in analytic form via a saddle-point integration. In addition, we have used the quench-action approach to obtain exact results for the stationary values of correlation functions in the XXZ chain at long times after the quench.
	
	These analytical results then formed the basis for our analysis of numerical data for the quench dynamics in the XXZ chain obtained by the light-cone renormalization group, a variant of the density-matrix renormalization group. Quite remarkably, we find that the XZ asymptotics describes the numerical data extremely well for $\Delta>3$ if a shift in the stationary value---exactly known from the QA---and a small renormalization of the amplitudes is taken into account. We find, in particular, that the power-law exponents of the decay towards the stationary value as well as the oscillation frequencies remain unchanged. 
	
	In summary, we find that the behavior of correlators $\langle\Psi_N|\sigma^z_0(t)\sigma^z_\ell(t)|\Psi_N\rangle$ at long times $t$ after the quench is described by an oscillatory behavior with oscillation amplitudes which decay as a power law with interaction-independent exponents and a leading power $t^{-3/2}$. We note that in contrast, the stationary values, oscillation frequencies, and amplitudes of the oscillations all do depend on the anisotropy $\Delta$ of the quench Hamiltonian. 
	
	For the future, it would be interesting to extend this study to quenches where the anisotropy of the quench Hamiltonian is within the critical phase of the XXZ chain. If we start from the N\'eel state this will be, however, never a small quench and there seems to be no effective model which could potentially aid in the analysis of the numerical data. Furthermore, the entanglement entropy in the numerical simulations does grow much more rapidly in this case, further limiting the obtainable simulation times.

\begin{acknowledgments}
This work was supported by the Deutsche Forschungsgemeinschaft (DFG) via the SFB/Transregio 185 (A4) and the research unit FOR 2316. The authors thank Compute Canada and the high performance cluster “Elwetritsch”  for providing computational resources. J.S. acknowledges support by the Natural Sciences and Engineering Research Council (NSERC) of Canada and thanks F.~Essler and J.-S.~Caux for discussions.
\end{acknowledgments}

	\appendix
	\section{Asymptotic formulas for large-$\Delta$ limit } \label{asymapp}
	
	As mentioned in the main text, to obtain the asymptotics of distance-$\ell$ correlators we must calculate the expectation value of an operator product [see Eq.(\ref{corrma})] that can be written as a Pfaffian of pairwise contractions. The problem is then reduced to calculating two-point correlations. In particular, for  values of $\ell$ up to 3, the contractions involved are $\bra A_0B_1 \ket $, $\bra A_2B_1\ket$, and $\bra A_0B_3\ket$. In this appendix, we provide some details about the calculation of such contractions.
	
	The integral form of $\bra A_mB_n\ket$ is shown in Eq.~\eqref{ambn}. Let us first focus on the stationary values given by 
	\be
	\bra A_mB_n\ket_\infty=\int_{-\pi}^{\pi}\frac{dk}{2\pi}e^{-i[k(m-n)+2\theta_k]}\cos(2\theta_k+k).\label{statxz}
	\ee
	Using Euler's formula, trigonometric relations and observing that $\cos(2\theta_k)=2(2+\Delta)\cos(k)/\varepsilon_k$, one can write Eq.~(\ref{statxz}) as
	\begin{eqnarray}
	\bra A_mB_n\ket_\infty&=&\int_{-\pi}^{\pi}\frac{2dk}{\pi\varepsilon_k^2}\left\{\left\{2\cos[k(r+1)]+\Delta\cos[k(r-1)]\right\}\right.\nonumber\\
	&\times&\left.\left[2\cos(2k)+\Delta\right]\right\},\label{asinf}
	\end{eqnarray}
	where $r=n-m$. Thus, the stationary values of the contractions $\bra A_0B_1 \ket $, $\bra A_2B_1\ket$, and $\bra A_0B_3\ket$ are
	\begin{eqnarray}
	\bra A_0B_1 \ket_\infty&=&\frac{\Delta^2-2}{\Delta^2},\\
	\bra A_2B_1 \ket_\infty&=&\frac{1}{\Delta},\\
	\bra A_0B_3 \ket_\infty&=&\frac{4-\Delta^2}{\Delta^3}.
	\end{eqnarray}  
	
	Now, we turn to the time-dependent contributions given by
	\begin{eqnarray}
	\bra A_mB_n \ket_t=i\int_{-\pi}^{\pi}\frac{dk}{2\pi}&&\left\{e^{-i[k(m-n)+2\theta_k]}\cos(2\varepsilon_k t)\right.\nonumber\\
	&\times&\left.\sin(2\theta_k+k)\right\}.\label{abt}
	\end{eqnarray}
	It is convenient to write Eq. (\ref{abt}) in the following form
	\begin{eqnarray}
	\bra A_mB_n \ket_t&=&\int_{-\frac{\pi}{2}}^{\frac{\pi}{2}}\frac{4dk}{\pi\varepsilon_k^2}\left\{e^{i2\varepsilon_kt}\{2\sin[k(r+1)]\right.\nonumber\\&+&\left.\Delta\sin[k(r-1)] \}\sin(2k) + \text{h.c.}\right\}.
	\end{eqnarray}
	
	To obtain the asymptotics of $\bra A_mB_n \ket_t$, we carry out a saddle-point analysis. The leading terms in the long-time regime are determined by the stationary points of $\varepsilon_k$, which are $k_0=0$ and  $\pm\pi/2$. Expanding the dispersion relation in the vicinity of $k_0=0$ and  $\pm\pi/2$ up to fourth order, we respectively have
	\begin{eqnarray}
	\varepsilon_k &\approx&  4\left(\frac{\Delta}{2}+1\right)-\frac{8\Delta}{\Delta+2}k^2+\frac{8\Delta(\Delta^2-2\Delta+4)}{3(\Delta+2)^3}k^4, \nonumber \\
	\varepsilon_k&\approx& 4\left(\frac{\Delta}{2}-1\right)+\frac{8\Delta}{\Delta-2}\left(k\pm \frac{\pi}{2}\right)^2 \nonumber\\
	&-&\frac{8\Delta(\Delta^2+2\Delta+4)}{3(\Delta-2)^3}\left(k\pm\frac{\pi}{2}\right)^4.
	\end{eqnarray}
	Assuming $t\gg1$ and denoting $x^2=k^2t$, we can write $\bra A_mB_n\ket_t$ for $k_0=0$ as 
	\begin{widetext}
	\begin{eqnarray}
\bra A_mB_n\ket_{t,0}&=&\frac{32e^{8i(1+\Delta/2)t}\xi(r,\Delta)}{\sqrt\pi t^{3/2}}\left(1+\frac{2}{\Delta}\right)^{-1/2}\int_{-\infty}^{\infty}dx e^{-16i\Delta x^2/(\Delta+2)}\left\{  2x^2-\frac{1}{t}\left[\frac{2(r+1)^3+\Delta(r-1)^3}{6(r+1)+3\Delta(r-1)}\right.\right.\nonumber\\
&+& \left.\left.\frac{4(4-8\Delta+\Delta^2)}{3(2+\Delta)^2} \right]x^4+\frac{32 i \Delta (4-2\Delta+\Delta^2)}{3(2+\Delta)^3}x^6\right\}+\text{h.c.} +O(t^{-7/2}),\label{saddint}
\end{eqnarray}
	\end{widetext}
with $\xi(r,\Delta)$ 
\begin{equation}
	\xi(r,\Delta)=\frac{2(r+1)+\Delta(r-1)}{128\sqrt\pi(1+\Delta^2/4+\Delta)}\left(1+\frac{2}{\Delta}\right)^{1/2} .
\end{equation}

	By a deformation of the complex contour, the integral in Eq.~(\ref{saddint}) can be solved by transforming it into an exponentially decaying one. Doing so, we have
	\begin{eqnarray}
\bra A_mB_n\ket_{t,0}&\approx&\xi(r,\Delta) 
\left\{\left(1+\frac{2}{\Delta}\right)\frac{\sin\left[8\left(\frac{\Delta}{2}+1\right)t-\frac{\pi}{4}\right]}{t^{3/2}}\right.\nonumber\\
 &+&\left.
 \frac{\chi(r,\Delta)}{16\Delta^2}\frac{\cos\left[8\left(\frac{\Delta}{2}+1\right)t-\frac{\pi}{4}\right]}{t^{5/2}}\right\},\label{as0}
\end{eqnarray}
where the prefactor $\chi(\Delta)$ is given by
	\begin{eqnarray}
	\chi(r,\Delta)&=&\left(1+\frac{\Delta^2}{4}\right)\left[\frac{2(r+1)^3+\Delta(r-1)^3}{2(r+1)+\Delta(r-1)}-1\right]\nonumber\\&+& \Delta\left[\frac{2(r+1)^3+\Delta(r-1)^3}{2(r+1)+\Delta(r-1)}-\frac{11}{2}\right]. 
	\end{eqnarray}
	The saddle points at $k_0=\pm\pi/2$ give contributions to the asymptotic limit which can be obtained in an analogous manner 
	to those for $k_0=0$. For odd values of $|r|$, we find 
	\begin{eqnarray}
	&&\bra A_mB_n\ket_{t,\pi/2}\approx\frac{\xi(r,-\Delta)}{2} 
	\left\{\left(\frac{2}{\Delta}-1\right)\right.\\
	&\times&\left.\frac{\sin\left[8\left(\frac{\Delta}{2}-1\right)t-\frac{3\pi}{4}\right]}{t^{3/2}}\right.\nonumber\\
	&+&
	\frac{\chi(r,-\Delta)}{16\Delta^2} \left.\frac{\cos\left[8\left(\frac{\Delta}{2}-1\right)t-\frac{3\pi}{4}\right]}{t^{5/2}}\right\}(-1)^{(r+1)/2}.\label{aspi} \nonumber
	\end{eqnarray}
	Finally, the complete expression for the contraction $\bra A_mB_n\ket$ is given by
	\begin{equation}
	\bra A_mB_n \ket= \bra A_mB_n \ket_\infty+\bra A_mB_n \ket_{t,0}+2\bra A_mB_n \ket_{t,\pi/2}.\label{fullcont}
	\end{equation}
	
\section{The quench action approach}
\label{QAA}
Here, we provide further explanations about the calculation of the asymptotic values of correlation functions in the long-time limit using the quench-action approach.  The procedure involves solving the generalized TBA (gTBA) integral equation, whose driving term is determined by the overlap of the initial state with the quench state. This driving term determines the saddle-point $\eta$-functions through the gTBA formula, which with two additional auxiliary functions then determine the long-time limit of the equal-time correlations along the $z$-direction and more generally any axis. Here, the relevant formulas are collected, for additional details about the derivation see Refs.~\cite{Brockmann_2014,mestyan_2015}.

For clarity, the different kernels that will be used are collected below  in approximate order of appearance. First, the kernels necessary to determine the saddle-point $\eta$ functions and relevant auxiliary functions are given by 
\begin{eqnarray}
    s(\lambda) &=& \frac{1}{2\pi } \sum_{k \in \mathbb{Z}} \frac{ e^{-2 i k \lambda}}{\cosh(k \eta)} = \frac{1}{2\pi }\left(1 +  2\sum_{k =1} \frac{ \cos{2 k \lambda}}{\cosh(k \eta)} \right), \nonumber \\
    t(\lambda) &=& \frac{1}{2\pi } \sum_{k=1}^\infty \frac{\sinh(k \eta)}{\cosh^2(k \eta)} \sin(2 k \lambda) \, .
\end{eqnarray}
We will use the shorthand notation for the derivatives of these kernels
\begin{eqnarray}
    s^{(j)}(\lambda) &=& \partial_\lambda^j s(\lambda), \nonumber \\
    t^{(j)}(\lambda) &=& \partial_\lambda^j t(\lambda).
\end{eqnarray}
A second set of kernels, necessary for computing correlations, are 
\begin{eqnarray}
a_n(\lambda) = \frac{i}{2\pi } \frac {\partial} {\partial \lambda} \ln \left( \frac{\sin(\lambda + i n \eta /2)}{\sin(\lambda - i n \eta /2)} \right),\\
b_n(\lambda) = \frac{i}{2\pi } \frac {\partial} {\partial \eta} \ln \left( \frac{\sin(\lambda + i n \eta /2)}{\sin(\lambda - i n \eta /2)} \right),
\end{eqnarray}
with the derivatives of these kernels again denoted by the shorthand
\begin{eqnarray}
    a_n^{(j)}(\lambda) = \partial_\lambda^j a_n(\lambda), \nonumber \\
    b_n^{(j)}(\lambda) = \partial_\lambda^j b_n(\lambda) \, .
\end{eqnarray}
Finally, the driving term for a N\'eel to XXZ quench is given in terms of Jacobi theta-functions with nome $q^2=\text{e}^{-2\eta}$
\begin{eqnarray}
    d_n(\lambda)  = (-1)^n \ln \left[ \left(\frac{\theta_4(\lambda)}{\theta_1(\lambda)} \right)^2\right] + \ln \left[ \left(\frac{\theta_2(\lambda)}{\theta_3(\lambda)} \right)^2\right]\, .\nonumber \\
\end{eqnarray}

With these kernels and the driving term now defined, the gTBA equation for $\cosh(\eta)= (q+q^{-1})/2=\Delta > 1$ can be evaluated. It is particularly convenient to consider the gTBA in its uncoupled form, which determines the $\eta_n$ functions recursively as
\begin{eqnarray}
    \ln \eta_n = d_n + s \star \left[ \ln (1+\eta_{n-1}) + \ln(1 + \eta_{n+1})\right],
\end{eqnarray}
with the shorthand $ s \star g = \int_{-\pi/2}^{\pi/2} d\mu \, s(\lambda - \mu) g(\mu)$. This formula is combined with both $\eta_0 =1$ and secondly, for large $n$,
\begin{eqnarray}
    \lim_{n \to \infty} \eta_{2n}(\lambda)&=& \eta_{\text{even}}(\lambda) \, , \\
        \lim_{n \to \infty} \eta_{2n+1}(\lambda)&=& \eta_{\text{odd}}(\lambda).
\end{eqnarray}
Often, these large $n$ formulas are satisfied very quickly, especially for large values of $\Delta $.

These saddle-point functions can be used to solve two additional integral relations. The first characterizing $\rho^{(j)}$ is given by
\begin{eqnarray}
    \rho_n^{(j)}(\lambda) = \delta_{n1} s^{(j)}(\lambda) + \left[ s \star \left( \frac{\rho_{n-1}^{(j)}}{1 + 1/\eta_{n-1}} + \frac{\rho_{n+1}^{(j)}}{1 + 1/\eta_{n+1}}\right) \right],\nonumber\\
\end{eqnarray}
with $\rho^{(j)}_0 =0$, in general. The second set of auxiliary functions $\sigma^{(j)}$ are determined by
\begin{eqnarray}
    \sigma_n^{(j)}(\lambda) &=& \delta_{n1} t^{(j)}(\lambda) + \left[ t \star \left( \frac{\rho_{n-1}^{(j)}}{1 + 1/\eta_{n-1}} + \frac{\rho_{n+1}^{(j)}}{1 + 1/\eta_{n+1}}\right) \right] \nonumber\\
    &&+\left[ s \star \left( \frac{\sigma^{(j)}_{n-1}}{1 + 1/\eta_{n-1}} + \frac{\sigma^{(j)}_{n+1}}{1 + 1/\eta_{n+1}}\right) \right],
\end{eqnarray}
with $\sigma^{(j)}_0=0$ similarly to the previous case. As in the case of the saddle-point $\eta$-functions at large values of $n$, the even and odd functions converge to two common functions, allowing us to truncate the series.

These auxiliary functions are now used to determine the value of the two quantities
\begin{eqnarray}
\Gamma_{jk} &=& -4 \pi \sum_{n=1} \int_{-\pi/2}^{\pi/2}  d\lambda \left[ a^{(k)} \frac{ \sigma_n^{(j)}}{1 + 1/\eta_n} +b^{(k)} \frac{ \rho_n^{(j)}}{1 + 1/\eta_n} \right], \nonumber \\
    \Omega_{jk} &=& 4 \pi \sum_{n=1} \int_{-\pi/2}^{\pi/2} d\lambda \, a^{(k)} \frac{ \rho_n^{(j)}}{1 + 1/\eta_n} \, ,
\end{eqnarray}
which in combination with a final set of functions
\begin{eqnarray}
K(u) &=& \frac{\sinh(2\eta)}{\sinh(u+\eta) \sinh(u-\eta)}, \nonumber\\
\widetilde{K}(u) &=& \frac{\sinh(2 u)}{\sinh(u+\eta) \sinh(u-\eta)},
\end{eqnarray}
define the convenient variables
\begin{eqnarray}
\omega_{ab} &=& -(-1)^{(a+b)/2} \Omega_{ab} - (-1)^b\frac{1}{2} \left(\frac{\partial}{\partial u}\right)^{a+b} K(u) \bigg|_{u=0}, \nonumber\\
W_{ab} &=& (-1)^{(a+b+1)/2} \Gamma_{ab} + (-1)^b\frac{1}{2} \left(\frac{\partial}{\partial u}\right)^{a+b} \widetilde{K}(u) \bigg|_{u=0} \, . \nonumber \\
\end{eqnarray}
Finally, these quantities may be inserted into QTM identities for the relevant stationary correlations. The explicit relations for determining the relevant correlations considered in the main text are reported below: 
\begin{widetext}
\begin{eqnarray}
C(1,\infty) &=&\langle \sigma_1^z \sigma_2 ^z \rangle = \coth(\eta) \omega_{00} + W_{10},\\
C(2,\infty) &=& \langle \sigma_1 ^z \sigma_3^z \rangle = 2 \coth(2\eta) \omega_{00} + W_{10} +\frac{ \omega_{20} - 2 \omega_{11}}{4} \tanh(\eta) - \frac{\sinh^2(\eta)}{4} W_{21},\\
C(3,\infty) &=& \langle \sigma_1^z \sigma^z_4 \rangle = \frac{4}{768 \sinh(4\eta) \cosh(2\eta)} \bigg[ 384q^4 (1+q^2)^2 ( 5 - 4 q^2 + 5 q^4) \omega_{00} \nonumber \\
&&-8 (1+ q^4( 52 + 64q^2 - 234q^4 + 64q^6 + 52 q^8 + q^{12})) \omega_{11} +192 q^4 (q^2 - 1)^2 (1 + 4q^2 + q^4) \omega_{02}\nonumber \\
&&+(q^2 -1)^4 (1+q^4) (1+4 q^2 + q^4)(-4 \omega_{13} + 6 \omega_{22}) -768 q^4 (-1 - q^2 + q^6+q^8) W_{01}\nonumber\\
&&+16(q^2-1)^3(1+6q^2+11q^4+11 q^6+6 q^8 +q^{10}) W_{12}-2(q^2-1)^5 (1+2 q^2+2 q^4 + q^6) W_{23}\nonumber\\
&&+ 8(q^2 -1)^3 (1+q^2)(1+6q^2+34q^4+6q^6+q^8)(\omega_{01}^2 - \omega_{00} \omega_{11}))\nonumber\\
&&+(-1-4q^2-22 q^4 - 12 q^6 + 12 q^{10} + 22q^{12} + 4 q^{14}+q^{16}) (-6 \omega_{02}^2 + 12 \omega_{02} \omega_{11} + 4 \omega_{01} \omega_{12} - 4 \omega_{00} \omega_{13} + 6 \omega_{00} \omega_{22} )\nonumber\\
&&+16(q^2 - 1)^4 (1+q^2)^2 (1+q^2+q^4) (\omega_{02}W_{01} - \omega_{01} W_{02} + \omega_{00} W_{12})\nonumber\\
&&+(q^4-1)^2 (1+5q^2+6q^4 + 5q^6+q^8) (4 \omega_{13} W_{01} - 6 \omega_{22} W_{01} - 2 \omega_{03} W_{02} \nonumber\\
&&\,\,\,\,\,\, +6 \omega_{12} W_{02} +2 \omega_{02}W_{03} - 4 \omega_{11} W_{03} - 6 \omega_{02}W_{12} + 4 \omega_{01} W_{13} - 2 \omega_{00} W_{23})\nonumber\\
&&+3(q^4-1)^3 (1+q^2 + q^4) (W_{03}W_{12} - W_{02} W_{13} + W_{01} W_{23}) \bigg].
\end{eqnarray}
\end{widetext}

\end{document}